\documentclass[twocolumn,showpacs,preprintnumbers,amsmath,amssymb,superscriptaddress]{revtex4}

\usepackage{graphicx}
\usepackage{dcolumn}
\usepackage{bm}
\usepackage{color}

\begin{document}

\preprint{APS}

\title{Coexistence of localized and itinerant electrons in BaFe$_2 X_3$ 
($X$ = S and Se) revealed by photoemission spectroscopy}

\author{D.~Ootsuki}
\affiliation{Department of Physics, University of Tokyo, Kashiwa, Chiba 277-8561, Japan}
\author{N.~L.~Saini}
\affiliation{Department of Physics, Universit\'a di Roma "La Sapienza", Piazzale Aldo Moro 2, 00185 Roma, Italy}
\author{F.~Du}
\affiliation{Institute for Solid State Physics, University of Tokyo, Kashiwa, Chiba 277-8581, Japan}
\affiliation{Key Laboratory of Physics and Technology for Advanced Batteries (Ministry of Education), College of Physics, 
Jilin University, Changchun, 130012, People's Republic of China}
\author{Y.~Hirata}
\affiliation{Institute for Solid State Physics, University of Tokyo, Kashiwa, Chiba 277-8581, Japan}
\author{K.~Ohgushi}
\affiliation{Department of Physics, Tohoku University, Sendai, Miyagi 980-8578, Japan}
\affiliation{Institute for Solid State Physics, University of Tokyo, Kashiwa, Chiba 277-8581, Japan}
\author{Y.~Ueda}
\affiliation{Toyota Physical and Chemical Institute, Nagakute, Aichi 480-1192, Japan}
\affiliation{Institute for Solid State Physics, University of Tokyo, Kashiwa, Chiba 277-8581, Japan}
\author{T.~Mizokawa}
\affiliation{Department of Complexity Science and Engineering, 
University of Tokyo, 5-1-5 Kashiwanoha, Kashiwa, Chiba 277-8561, Japan}
\affiliation{Department of Physics, University of Tokyo, Kashiwa, Chiba 277-8561, Japan}
\affiliation{Department of Physics, Universit\'a di Roma "La Sapienza", Piazzale Aldo Moro 2, 00185 Roma, Italy}

\date{\today}

\begin{abstract}
We report a photoemission study at room temperature 
on BaFe$_2 X_3$ ($X$ = S and Se) and CsFe$_2$Se$_3$ in which 
two-leg ladders are formed by the Fe sites.
The Fe 2$p$ core-level peaks of BaFe$_2 X_3$ are broad and exhibit 
two components, indicating that itinerant and localized 
Fe 3$d$ sites coexist similar to K$_x$Fe$_{2-y}$Se$_2$. 
The Fe 2$p$ core-level peak of CsFe$_2$Se$_3$ is rather sharp 
and is accompanied by a charge-transfer satellite. 
The insulating ground state of CsFe$_2$Se$_3$ can be viewed 
as a Fe$^{2+}$ Mott insulator in spite of the formal valence
of +2.5. The itinerant versus localized behaviors
can be associated with the stability of chalcogen $p$ holes 
in the two-leg ladder structure.
\end{abstract}

\pacs{74.25.Jb, 74.70.Xa, 79.60.-i, 74.81.-g}
\maketitle

\newpage

\section{Introduction}

The coexistence of itinerant superconducting phase and
localized antiferromagnetic phase in K$_x$Fe$_{2-y}$Se$_2$
\cite{Guo2010, Shermadini2011, Ye2011, Ricci2011}
sheds light on the intervening coupling between the electron
correlation effect and the lattice effect.
When the Fe 3$d$ electrons are localized and form the antiferromagnetic
insulating state with high-spin Fe$^{2+}$, the insulating phase
tends to expand due to Fe-Fe bond length increase and applies 
a kind of pressure to the remaining metallic region. \cite{Saini}
The antiferromagnetic insulating phase in K$_x$Fe$_{2-y}$Se$_2$
is identified as K$_2$Fe$_4$Se$_5$ with Fe vacancy order.
\cite{Shermadini2011, Ye2011}
On the other hand, the superconducting phase is likely to be FeSe 
which is under the pressure from the expanded and insulating K$_2$Fe$_4$Se$_5$.
In the insulating K$_2$Fe$_4$Se$_5$ phase,
the four Fe sites with square geometry form a ferromagnetic block
which couples antiferromagnetically with neighboring blocks. \cite{Ye2011} 
The transfer integrals between the neighboring blocks 
are reduced due to the Fe vacancy, and the Mott insulating state is
realized. In the XPS study on K$_x$Fe$_{2-y}$Se$_2$, two components
of the Fe 2$p_{3/2}$ peak are assigned to the coexisting superconducting 
and insulating phases in the superconducting K$_x$Fe$_{2-y}$Se$_2$ while, 
in insulating K$_x$Fe$_{2-y}$Se$_2$, the Fe  2$p_{3/2}$ peak consists of 
a single component. \cite{Oiwake2013}
The charge-transfer energy from Se 4$p$ to Fe 3$d$ is estimated 
to be 2.3 eV which is smaller than the repulsive Coulomb interaction 
between the Fe 3$d$ electrons of 3.5 eV. \cite{Oiwake2013} 
Therefore, the insulating K$_2$Fe$_4$Se$_5$ phase with high-spin Fe$^{2+}$ 
can be regarded as a Mott insulating state of charge-transfer type
instead of Mott-Hubbard-type,
and the Se 4$p$ orbitals should be taken into account to explain
the magnetic interactions between the Fe spins.
 
Recently, another insulating Fe selenide BaFe$_2$Se$_3$ has been attracting 
much attention due to the specific quasi one-dimensional crystal structure 
with Pnma space group as well as the block-type antiferromagnetic state 
similar to K$_2$Fe$_4$Se$_5$.
\cite{Krzton-Maziopa2011, Caron2011, Lei2011, Caron2012, Nambu2012, Monney2013, Luo2013}
In addition, BaFe$_2$Se$_3$ is predicted to be a new type of multiferroic system 
with magnetic and ferrielectric orders in a recent theoretical work, \cite{Dong2014}
which certainly enhances the interest in the title system. 
In BaFe$_2$Se$_3$, FeSe$_4$ tetrahedra share their edges and form a two-leg ladder 
of Fe sites as shown in Fig. 1(a).
The magnetic structure of BaFe$_2$Se$_3$ is similar to K$_2$Fe$_4$Se$_5$
in that four Fe spins in the two-leg ladder form a ferromagnetic block 
and the neighboring blocks are antiferromagnetically coupled as illustrated 
in Fig. 1.
The two-leg ladder is distorted with shorter Fe-Fe bonds 
(ferromagnetic and antiferromagnetic) and longer Fe-Fe bonds
(ferromagnetic and antiferromagnetic) along the ladder direction.
The magnetic structure is consistent with the theoretical 
prediction. \cite{Medvedev2012}
N\'eel temperature reported in the literatures varies from 
240 K \cite{Krzton-Maziopa2011} to 255 K. \cite{Lei2011, Caron2011, Nambu2012}
Also the magnetic moment of BaFe$_2$Se$_3$ ranges from 
2.1 $\mu_B$ \cite{Krzton-Maziopa2011} to 2.8 $\mu_B$ \cite{Caron2011, Nambu2012} 
depending on the growth condition, and is smaller than the value expected 
for high-spin Fe$^{2+}$.
Interestingly, resistivity also depends on the growth condition.
Lei {\it et al.} reported that resistivity exhibits activation-type temperature 
dependence with band gap of 0.18 eV. \cite{Lei2011} On the other hand,
one-dimensional variable range hopping was reported by Nambu {\it et al.}
indicating that some carriers are localized due to strong scattering effect 
in the quasi one-dimensional structure. \cite{Nambu2012}
In addition, coexistence of itinerant and localized electrons
was recently indicated by the resonant inelastic x-ray scattering study 
by Monney {\it et al}. \cite{Monney2013}
This observation suggests that the itinerant electrons introduced by 
small Fe vacancy or some other effects would be responsible for 
the reduction of the magnetic moment and the variable range hopping 
behavior of the resistivity. \cite{Nambu2012}
In contrast to BaFe$_2$Se$_3$, CsFe$_2$Se$_3$ with formal Fe valence 
of +2.5 is much more insulating. \cite{Fei2012}
Interestingly, Fei {\it et al.} have revealed that all the Fe sites 
have the same local environment in CsFe$_2$Se$_3$ 
by means of M\"ossbauer spectroscopy. \cite{Fei2012}
Below 177 K, Fe spins in the two-leg ladder of CsFe$_2$Se$_3$ order
antiferromagnetically along the rung and ferromagnetically
along the leg with magnetic moment of $\sim$ 1.8 $\mu_B$.
Usually, Mott insulators with integer number of valence 
are expected to be more insulating than the mixed valence systems.
The situation of the two-leg ladder Fe chalcogenides 
is opposite to this expectation, and CsFe$_2$Se$_3$ with formal Fe
valence of +2.5 is more insulating than integer valence 
BaFe$_2$Se$_3$ and BaFe$_2$S$_3$. \cite{Fei2012}
Such puzzling mismatch between the formal valence 
and the transport behavior indicates unusual electronic states
of the title system.

\begin{figure}
\includegraphics[width=0.5\textwidth]{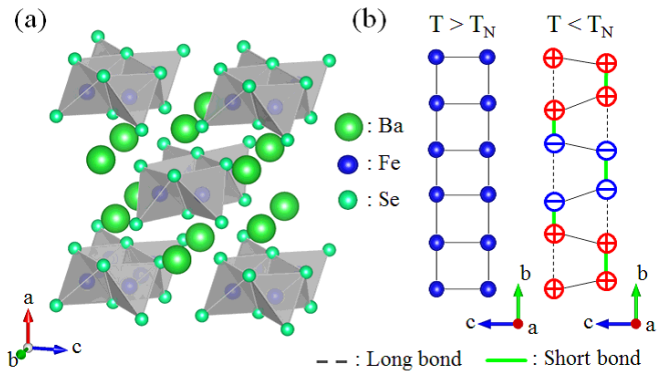}
\caption{
(Color online) (a) Crystal structure of BaFe$_2$Se$_3$ visualized using the software package Vesta\cite{vesta}. 
(b) Schematic drawing for the magnetic structure and the lattice distortion
for BaFe$_2$Se$_3$.
}
\end{figure}

In the present work, we study fundamental electronic structures 
of BaFe$_2$Se$_3$, BaFe$_2$S$_3$, and CsFe$_2$Se$_3$ above their 
N\'eel temperatures by means of x-ray photoemission spectroscopy (XPS) and
ultra-violet photoemission spectroscopy(UPS) at room temperature.
The broad Fe 2$p$ XPS peaks of BaFe$_2$Se$_3$ and BaFe$_2$S$_3$ 
indicate coexistence of localized and itinerant electrons.
On the other hand, the Fe 2$p$ XPS peak of CsFe$_2$Se$_3$ is relatively sharp 
suggesting that Fe valence is homogeneous in spite of the expectation of a mixed valence state.
The apparent contradiction between the valence state and the Fe 2$p$ peak width 
can be reconciled by taking account of the Se 4$p$ or S 3$p$ holes. 

\section{Experiments}

The single crystals of BaFe$_2$Se$_3$, BaFe$_2$S$_3$, and CsFe$_2$Se$_3$
were grown as reported in the literatures. \cite{Nambu2012, Fei2012}
We cleaved the single crystals at room temperature (300 K) 
under the ultrahigh vacuum for the XPS and UPS measurements.
The XPS measurement was carried out at 300 K using JPS9200 analyzer.
Mg K$\alpha$ (1253.6 eV) was used as an x-ray source. 
The total energy resolution was set to $\sim$ 1.0 eV. 
The binding energy was calibrated using the Au 4$f$ core level 
of the gold reference sample.
The UPS measurement was performed using SES100 analyzer
and a He I (21.2 eV) source. The total energy resolution was set to 
$\sim$ 30 meV and the Fermi level was determined using the Fermi egde
of the gold reference sample.

\begin{figure}
\includegraphics[width=0.5\textwidth]{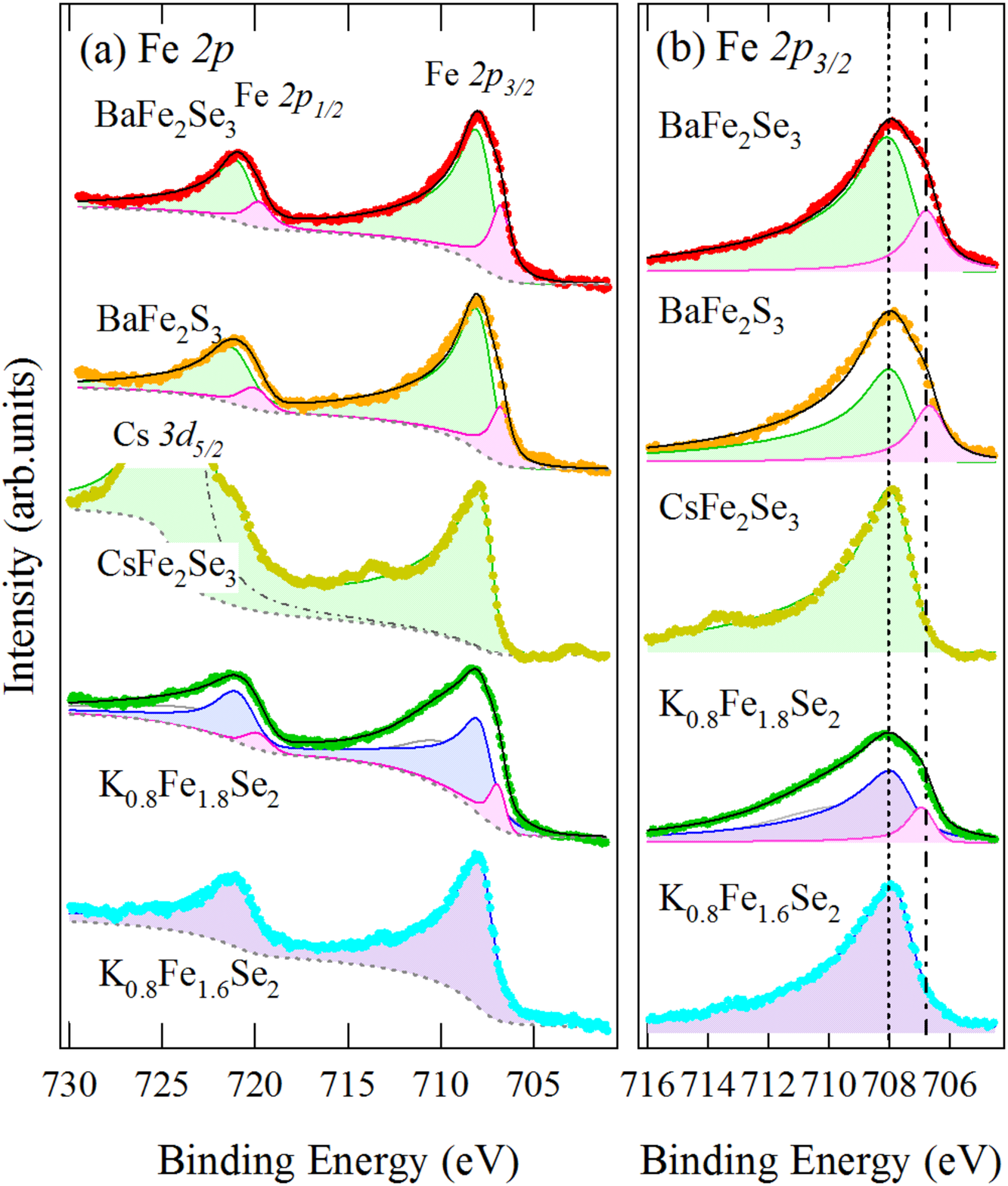}
\caption{
(Color online) Fe 2$p$ XPS of BaFe$_2 X_3$ ($X$ = S and Se) compared 
with CsFe$_2$Se$_3$ and K$_x$Fe$_{2-y}$Se$_2$ (metallic and insulating).
\cite{Oiwake2013} The solid curves indicate the results of curve fitting for BaFe$_2$S$_3$
and BaFe$_2$Se$_3$.
}
\end{figure}

\section{Results and Discussion}

Figure 2 shows the Fe 2$p$ XPS spectra of BaFe$_2$Se$_3$, BaFe$_2$S$_3$, 
and CsFe$_2$Se$_3$ taken at 300 K which are compared with those of 
superconducting and non-superconducting K$_x$Fe$_{2-y}$Se$_2$. \cite{Oiwake2013}
In non-superconducting K$_x$Fe$_{2-y}$Se$_2$, the chemical composition 
is close to K$_2$Fe$_{4}$Se$_5$ with Fe$^{2+}$ and its ground state 
is a charge-transfer-type Mott insulator. The sharp Fe 2$p$ peak of CsFe$_2$Se$_3$
is very similar to that of non-superconducting K$_x$Fe$_{2-y}$Se$_2$ or
the Fe$^{2+}$ Mott insulator. This XPS result indicates that 
CsFe$_2$Se$_3$ would be a Mott insulator with Fe$^{2+}$ 
which is actually consistent with the M\"ossbauer study. \cite{Fei2012}
If all the Fe sites in CsFe$_2$Se$_3$ take the high-spin Fe$^{2+}$ 
configuration, the extra positive charge (+0.5 per Fe) should be
located at the Se sites.
On the other hand, in superconducting K$_x$Fe$_{2-y}$Se$_2$,
itinerant Fe 3$d$ electrons coexist with the localized Fe 3$d$ electrons 
in the Mott insulating phase. The Fe 2$p$ lineshape of BaFe$_2$Se$_3$ 
and BaFe$_2$S$_3$ is very similar to that of superconducting
K$_x$Fe$_{2-y}$Se$_2$, indicating coexistence of itinerant and 
localized electrons. 
In case of superconducting K$_x$Fe$_{2-y}$Se$_2$, the coexistence
of the localized and itinerant electronic states is governed 
by the spatial distribution of the Fe vacancy.
On the other hand, BaFe$_2$Se$_3$ and BaFe$_2$S$_3$ have no Fe vacancy
which can reduce transfer integrals between neighboring Fe sites and cause
the Mott localization. Instead, in the two-leg ladder structure of BaFe$_2$Se$_3$ 
and BaFe$_2$S$_3$, the electronic interaction between neighboring Fe sites
can be controlled by the Se 4$p$ or S 3$p$ holes which are indicated 
by the mismatch between the formal valence and the transport behavior.
Here, one can speculate that the Fe 3$d$ electrons with 
the Fe$^{2+}$ high-spin configuration and the Se 4$p$ (or S 3$p$) holes
are localized in CsFe$_2$Se$_3$ whereas they are partially delocalized 
in BaFe$_2$Se$_3$ and BaFe$_2$S$_3$. 

\begin{figure}
\includegraphics[width=0.45\textwidth]{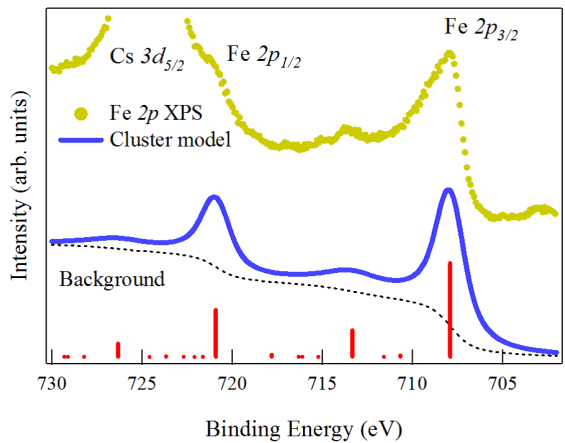}
\caption{
(Color online) Fe 2$p$ XPS of CsFe$_2$Se$_3$ (dots) compared with the result of cluster model calculation
The dotted curve indicates the background due to the secondary electrons. 
}
\end{figure}

The Fe 2$p_{3/2}$ and Fe 2$p_{1/2}$ peaks can be decomposed into the two components 
which are derived from the itinerant and localized parts.
In Fig. 2, the results of Mahan's line shape fitting 
are indicated by the solid curves for BaFe$_2$Se$_3$ and BaFe$_2$S$_3$.
The itinerant component has lower binding energy due to the 
stronger screening effect.
The intensity ratio between the itinerant and localized components
is 3.0 : 5.0 for BaFe$_2$Se$_3$ and 2.3 : 5.0 for BaFe$_2$S$_3$, respectively. 
The relative intensity of the ``itinerant'' component
is much larger in BaFe$_2$Se$_3$ and BaFe$_2$S$_3$ 
than that in K$_x$Fe$_{2-y}$Se$_2$.
The energy splitting between the two components is $\sim$ 0.8 eV 
for BaFe$_2$Se$_3$ and BaFe$_2$S$_3$ which is comparable to
the value for K$_x$Fe$_{2-y}$Se$_2$.

\begin{figure}
\includegraphics[width=0.47\textwidth]{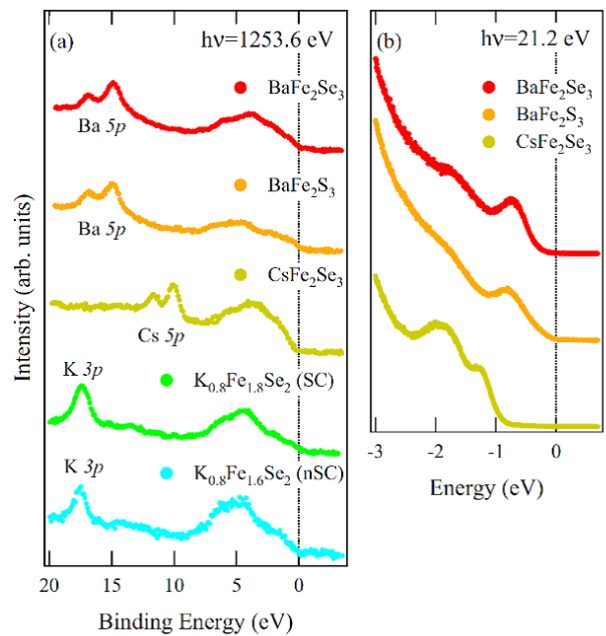}
\caption{
(Color online) (a) Valence-band XPS of BaFe$_2 X_3$  ($X$ = S and Se) 
compared with CsFe$_2$Se$_3$ and K$_x$Fe$_{2-y}$Se$_2$ (metallic and insulating).
\cite{Oiwake2013} (b) UPS of BaFe$_2 X_3$ ($X$ = S and Se) compared with CsFe$_2$Se$_3$.
}
\end{figure}

Figure 3 shows the Fe 2$p$ XPS of CsFe$_2$Se$_3$ compared with the result 
of the configuration interaction calculation on an FeSe$_4$ cluster model.
The Fe 2$p_{3/2}$ peak of CsFe$_2$Se$_3$ (located around 708 eV) 
is accompanied by a broad satellite structure (located around 714 eV)
which can be assigned to the charge-transfer satellite.
The energy position and the intensity of the charge-transfer satellite
can be analyzed using the configuration-interaction calculation 
on the tetrahedral FeSe$_4$ cluster model. \cite{Oiwake2013}  
The ground state is assumed to take the high-spin $d^6$ state 
mixed with the $d^7L$, $d^8L^2$, $d^9L^3$, and $d^{10}L^4$ states 
where $L$ represents a hole in the Se 4$p$ ligand orbitals.  
The excitation energy from $d^6$ to $d^7L$ corresponds to
the charge-transfer energy $\Delta$.
The excitation energy from $d^7L$ to $d^8L^2$ is given by $\Delta + U$
where $U$ represents the Coulomb interaction between the Fe 3$d$ electrons. 
The transfer integrals between the $d^nL^m$ and $d^{n+1}L^{m+1}$ are 
described by ($pd\sigma$) and ($pd\pi$),
where the ratio ($pd\sigma$)/($pd\pi$) is fixed at -2.16.
The final states are spanned by the $cd^6$, $cd^7L$, $cd^8L^2$, $cd^9L^3$, 
and $cd^{10}L^4$ states where $c$ denotes a hole of the Fe 2$p$ core level.
The Coulomb interaction $Q$ between the Fe 2$p$ hole and the Fe 3$d$ electron
is expressed as $Q$ which is fixed at $U/0.8$.
With $\Delta$ = 2.0 eV, $U$ = 3.5 eV, and ($pd\sigma)$ = -1.2 eV, the calculated spectrum 
can reproduce the Fe 2$p$ XPS result as indicated by the solid curve in Fig. 3. 
$\Delta$ is smaller than $U$, indicating that
CsFe$_2$Se$_3$ is a Mott insulator of charge-transfer typ
instead of Mott-Hubbard type.

The valence-band XPS spectra of BaFe$_2$Se$_3$, BaFe$_2$S$_3$, 
and CsFe$_2$Se$_3$ taken at 300 K are displayed in Fig. 4(a) 
and are compared with those of superconducting and 
non-superconducting K$_x$Fe$_{2-y}$Se$_2$. \cite{Oiwake2013}
Besides the shallow core levels such as Ba 5$p$, Cs 5$p$, and K 3$p$,
the valence-band structures of the five systems are similar to 
one another. The spectral weight near the Fermi level increases
in going from CsFe$_2$Se$_3$ to BaFe$_2$Se$_3$ to BaFe$_2$S$_3$,
consistent with their transport properties. \cite{Lei2011,Caron2012,Nambu2012,Fei2012}

Figure 4(b) shows the UPS spectra of BaFe$_2$Se$_3$, BaFe$_2$S$_3$, 
and CsFe$_2$Se$_3$ taken at 300 K. 
BaFe$_2$S$_3$ with the highest conductivity shows the tail of 
the spectral weight up to the Fermi level while BaFe$_2$Se$_3$ 
has the finite band gap of $\sim$ 0.2 eV.
The magnitude of the band gap observed in BaFe$_2$Se$_3$ 
is more or less consistent with the activation energies obtained 
from temperature dependence of resistivity by Lei {\it et al.} \cite{Lei2011}
CsFe$_2$Se$_3$ exhibits the largest band gap of $\sim$ 0.8 eV
consistent with the charge-transfer-type Mott insulator deduced
from the Fe 2$p$ XPS result.

The apparently homogeneous Fe valence and the relatively large band
gap in CsFe$_2$Se$_3$ can be explained based on the idea of ligand hole.
The smallness of the charge-transfer energy for the Fe$^{2+}$ state
indicates that, if Fe$^{3+}$ exists in CsFe$_2$Se$_3$, it should 
take the $d^6L$ configuration instead of $d^5$.
In this ligand hole picture, the two-leg ladder in CsFe$_2$Se$_3$
accommodates the $d^6$-like and $d^6L$-like sites. Assuming that
the $d^6$-like and $d^6L$-like sites are aligned along the rung,
the Se 4$p$ hole should be located at the Se sites sandwiched by
the two legs. This situation is schematically shown in Fig. 5
where all the Fe sites take the high-spin Fe$^{2+}$ configuration
and the S 4$p$ holes are localized at the Se sites on the rungs.
In this scenario, the Se 4$p$ holes and the Fe 3$d$ electrons 
are partially delocalized in BaFe$_2$Se$_3$ and BaFe$_2$S$_3$ 
and may cause the lattice (and magnetic) instability. 
In the low-temperature phase
with the lattice distortion and the block-type magnetic order, 
the Se 4$p$ holes and the Fe 3$d$ electrons would be localized
in the antiferromagnetic dimer with the Fe$^{2+}$ high-spin configuration, 
and they are rather ``itinerant''
in the ferromagnetic dimer. Here, the Se 4$p$ holes and 
the Fe 3$d$ electrons in the ferromagnetic dimer
are ``itinerant'' in a sense that they occupy a kind of molecular 
orbitals. 

\section{Conclusion}
\begin{figure}
\includegraphics[width=0.45\textwidth]{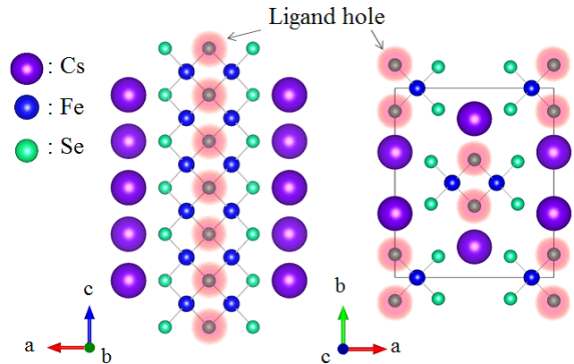}
\caption{
Crystal structure and possible Se 4$p$ hole distribution
for CsFe$_2$Se$_3$ visualized using the software package Vesta\cite{vesta}.
}
\end{figure}

In conclusion, we have studied the electronic structures of 
BaFe$_2 X_3$ ($X$ = S and Se) and CsFe$_2$Se$_3$ using 
photoemission spectroscopy. The Fe 2$p$ core-level peaks
consist of the two components in BaFe$_2 X_3$, indicating that 
the itinerant and localized Fe 3$d$ electrons coexist.
The Fe 2$p$ and valence-band spectra suggest that 
the itinerant Fe 3$d$ electrons are more strongly
confined in BaFe$_2$Se$_3$ than in BaFe$_2$S$_3$.
On the other hand, the Fe 2$p$ core-level peak of CsFe$_2$Se$_3$
exhibit the single component accompanied with the charge-transfer 
satellite. The insulating ground state of CsFe$_2$Se$_3$ can be 
viewed as a charge-transfer-type Mott insulator with 
localized Se 4$p$ holes. In BaFe$_2 X_3$ ($X$ = S and Se),
the Se 4$p$ or S 3$p$ holes are partially delocalized and 
may cause the coexistence of the itinerant and localized Fe 3$d$ electrons.
In future, the relationship between the chalcogen $p$ holes, 
the lattice distortions, and the magnetic interactions should be further studied 
using various experimental and theoretical approaches.

\section*{Acknowledgements}

The authors would like to thank Profs. H. Takahashi, H. Okamura, 
and Y. Uwatoko for valuable discussions.
This work was partially supported by Grants-in-Aid from the Japan Society of 
the Promotion of Science (JSPS) (Grant No: 25400356).
D.O. acknowledges supports from the JSPS Research Fellowship for Young Scientists.

\end{document}